\documentclass[11pt,final,onecolumn]{IEEEtran}
  
\usepackage{bm}
\usepackage{balance}
\usepackage{times}
\usepackage{amsmath,dsfont}
\usepackage{amssymb}
\usepackage{epsfig,verbatim} 
\usepackage{setspace}
\usepackage{color}
\usepackage{cite}
\usepackage{epstopdf}
\usepackage{subfigure}
\usepackage{graphics}
\usepackage{accents}
\usepackage{acronym}
\usepackage[bookmarks,colorlinks]{hyperref}
\usepackage{ulem}
\usepackage{mathtools}
\usepackage{algorithm}
\usepackage{algorithmic}
\usepackage{enumitem}
\usepackage[most]{tcolorbox}
\usepackage{cuted}
\usepackage{multicol,lipsum}

\title{In-Band Full-Duplex MIMO Systems\\for Simultaneous Communications and Sensing: Challenges, Methods, and Future Perspectives}

\author{Besma Smida,~\IEEEmembership{Senior~Member,~IEEE}, George C. Alexandropoulos,~\IEEEmembership{Senior~Member,~IEEE},\\ Taneli Riihonen,~\IEEEmembership{Senior~Member,~IEEE}, and Md Atiqul Islam,~\IEEEmembership{Member,~IEEE}
\thanks{B. Smida is with the Department of Electrical and Computer Engineering, University of Illinois Chicago, Chicago, IL 60601, USA. G. C. Alexandropoulos is with the Department of Informatics and Telecommunications, National and Kapodistrian University of Athens, Panepistimiopolis Ilissia, 15784 Athens, Greece and with the Department of Electrical and Computer Engineering, University of Illinois Chicago, Chicago, IL 60601, USA. T. Riihonen is with the Unit of Electrical Engineering, Tampere University, Korkeakoulunkatu~1, 33720 Tampere, Finland. M. A. Islam is with the Qualcomm Technologies, Inc., Santa Clara, CA 95051, USA.}
	\thanks{The work by B. Smida was in part supported by NSF CAREER number 1620902 and NSF CIF Medium number 1900911. The work by G. C. Alexandropoulos has been supported by the the Smart Networks and Services Joint Undertaking (SNS JU) 6G-DISAC project under the European Union's Horizon Europe research and innovation programme under Grant Agreement number 101139130.}}

\begin{document}
\maketitle

%%
%----------------------------------------------------------------------------------------
%	Abstract
%----------------------------------------------------------------------------------------
\begin{abstract}
\label{sec:abstract}
In-band Full-Duplex (FD) Multiple-Input Multiple-Output (MIMO) systems offer a significant  opportunity for Integrated Sensing and Communications (ISAC) due to their capability to realize simultaneous signal transmissions and receptions. This feature has been recently exploited to devise {\it spectrum-efficient} simultaneous information transmission and {\it monostatic} sensing operations, a line of research typically referred to as MIMO FD-ISAC. In this article, capitalizing on a recent FD MIMO architecture with reduced complexity analog cancellation, we present an FD-enabled framework for simultaneous communications and sensing using data signals. In contrast to communications applications, the framework's goal is not to mitigate self interference, since it includes reflections of the downlink data transmissions from targets in the FD node's vicinity, but to optimize the system parameters for the intended dual functionality. The unique characteristics and challenges of a generic MIMO FD-ISAC system are discussed along with a broad overview of state-of-the-art special cases, including numerical investigations. Several directions for future work on FD-enabled ISAC relevant to signal processing communities are also provided.
\end{abstract}
%, illustrating this finding with a running example of beamforming in multi-antenna ISAC systems
\acresetall

%%
%----------------------------------------------------------------------------------------
%	Introduction
%----------------------------------------------------------------------------------------
\section{Introduction }\label{sec:introduction}

%\textbf{Deadline --  After Mid-February or End-January.}  
Efficiently unifying sensing and communications --- two similar, yet so different and traditionally separated, wireless systems --- is of critical importance, gaining substantial attention in the current 3GPP studies~\cite{3GPPTS22137}. The sensing functionality, and the corresponding ability of the network to collect sensory data from the environment, give rise to new highly desirable capabilities in next generation wireless communication networks. To this end, the emerging Integrated Sensing and Communication (ISAC) paradigm allows for the exploitation of communication infrastructures in order to construct an environment-aware network~\cite{ETSIISG}. The ISAC technologies can considerably improve spectral and energy efficiency, while reducing both hardware and signaling costs, since they pursue to merge into a single system sensing and communication, which previously competed over various types of resources. However, %to date most proposed ISAC systems share the spectrum between radar and communication with minimum or no interference ~\cite{FMC22}. 
to date, most proposed integrated systems focus on \textit{coexistence}, i.e., \textit{non-overlapped resource allocation.}  They schedule these signals sequentially in the time domain or separate them in the frequency domain, or use spread-spectrum techniques, such as direct-spreading and time-hopping~\cite{FMC22}. It is well known that such approaches exhibit reduced spectrum efficiency.

\subsection{Motivation --- The moment is opportune for in-band full-duplex}
Capitalizing on recent achievements on In-Band Full-Duplex (IBFD) communication, we propose here simultaneously sensing target parameters and data communications over the same frequency bandwidth. Such FD-ISAC aims to increase the spectrum efficiency and reduce the latency of ISAC \cite{AlexandropoulosVTM_2023,10158711}. However, the coupling between the transmit and receive signals of an IBFD radio generates high-power Self-Interference (SI) at the Receiver (RX) side, which shadows the reflected signal from a remote target(s). As a consequence, the main challenge when implementing an IBFD radio is the SI signal which can be up to 100 dB stronger than the received signals of interest \cite{7756408}. Several works have provided experimental evidence and methods on SI suppression --- many are listed in the recent reviews~\cite{10463523,SmidaJSAC_2023}. Between 2010 and 2020, extensive research focused on IBFD wireless communication, leading to the integration of IBFD wireless products into 5G+ systems. This decade saw IBFD communication evolve from a concept in labs to a standard feature in telecommunications technology and products~\cite{SmidaJSAC_2023}.

\subsection{Scope --- Millimeter-wave massive MIMO FD-ISAC}
In this paper, we first outline {\it concisely} the basics of SI  in the context of IBFD Multiple-Input Multiple-Output (MIMO) systems. Our presentation on this interference entity and methods for its suppression are intentionally short since there are excellent reviews and tutorials available on the topic (e.g., \cite{SmidaJSAC_2023} and references therein). Then, we dive deeper into the innovation  related to  {\it massive} MIMO FD-ISAC, %ADD HERE -- 
%\GAadd{Our focus is monostatic sensing and we can present three separate scenarios with the respective optimization formulations: \textit{i}) communications' optimization and opportunistic sensing; \textit{ii}) sensing optimization and opportunistic communications; and \textit{iii}) joint optimization of both operations} 
which constitutes one of the most promising emerging technologies for realizing ISAC systems~\cite{FMC22}, combining IBFD and massive MIMO~\cite{AlexandropoulosVTM_2023} to realize simultaneous transmissions and receptions. The key benefit of devoting a large number of degrees-of-freedom (i.e., proportional to the number of antennas) per user is making beamforming more flexible. %  the large numbers ofantennas () can be used for beamforming to
%jointly suppress self-interference.% it is preferable for massive MIMO systems to operate uplink and downlink in the same spectrum, which is a requirement for IBFD. 
%Challenge: Self-interference cancellation techniques do not efficiently scale to massive MIMO base-stations with a large number of antennas.
%In massive MIMO systems, the base stations are equipped with a large number of antennas, ranging from 10s-100s, for simultaneous transmission to a small number of users. The key benefit of devoting a large number of degrees-of-freedom per user makes network management much easier while allowing a frequency reuse of one due to reduced inter-cell interference. 
However, SI reduction techniques developed for a small number of antennas do not efficiently scale to massive MIMO FD terminals with a large number of
antennas. This challenge gives rise to digital precoding solutions that are discussed in Section \ref{SIC}.

Concurrent to the massive MIMO developments, another
important trend has taken hold in the last decade. %To address the challenge of limited available spectrum in the sub-6 GHz bands, there has been a move to higher bands, especially the mmWave spectru
A major push to adopt millimeter-Wave (mmWave), and beyond, frequencies into cellular and WiFi standards, especially due to the availability of large unused spectrum therein. 
%Propagation studies demonstrating practical feasibility of mmWave-based networks [19] were instrumental in catalyzing that move. 
%For practical coverage, the use of large antenna arrays is critical in mmWave bands.
However, massive MIMO combined with wide bandwidths
poses a major challenge for achieving power-efficient Digital-to-Analog Conversion (DAC) and Analog-to-Digital Conversion (ADC). 
%Thus, all-digital beamforming is not practically feasible in mmWave bands, and hence hybrid beamformers are adopted. 
%Hybrid beamforming further constrains the precoding and combining opportunities available in sub-6 GHz massive MIMO systems, which have much smaller bandwidths. 
This unique set of challenges spiked a significant body of research in enabling mmWave massive MIMO FD-ISAC based on hybrid beamforming. 
%However, one of their main design challenges is the Self-Interference (SI) signal that is much stronger than the intended signal at the RX part of the FD node, which needs to be mitigated with SI cancellation techniques. ADD MASSIVE MIMO and WIDEBAND and COMPLEXITY and ARCHI. 
%--- PUT THIS SOMEWHERE ELSE
%Two reduced complexity hardware architectures for the analog canceler in FD MIMO comprising of reduced number of cancellation elements, compared to the state of the art, and simple multiplexers for efficient signal routing among the transceiver radio-frequency chains were presented in~\cite{FD_MIMO_Arch}. The one architecture was based on analog taps and the other on AUXiliary (AUX) Transmitters (TXs). In contrast to the available analog cancellation architectures, the values for each tap or each AUX TX and the configuration of the multiplexers were jointly designed with the digital transceiver beamforming filters according to desired performance objectives.  
%The former tap-based analog canceler architecture for SI cancellation was extended to the wideband case in~\cite{WB_FD_MIMO_TWC2022}, where also TX IQ imbalances and power amplification nonlinearities were handled via a joint optimization of Analog and Digital (A/D) SI cancellation together with the TX and receiver (RX) digital beamforming. 
In \cite{barneto2020beamforming}, a multi-beam FD-ISAC system was presented, according to which the Transmitter (TX) and RX beamformers were optimized to have multiple beams for both communications and sensing. In \cite{liyanaarachchi2021joint}, the RX spatial signal was further used to estimate the range and angle profiles corresponding to targets. In \cite{Comms-Target_Tracking2022,ISAC2022,FD_HMIMO_2023,Asilomar_FD_HMIMO_2023}, FD-ISAC systems including a massive MIMO TX, equipped with hybrid Analog and Digital (A/D) beamformers, communicating with multiple Downlink (DL) users while simultaneously estimating, via the same signaling waveforms, parameters related to radar targets were presented for both the near- and far-field regimes. In~\cite{Direction-Aided2020}, a direction-assisted beam management framework, where the TX was equipped with a large antenna array realizing DL analog
beamforming while few digitally-controlled reception antenna elements were used for estimating the Direction-of-Arrival (DoA) of the Uplink (UL) signal from an intended user, was presented.

%The concept of FD-ISAC has a long history in the context of radars

%\cite{AlexandropoulosVTM_2023,SmidaJSAC_2023} 
%-------------------------------------------------------------------------
%	Section I: FD MIMO Fundamentals
%-------------------------------------------------------------------------

\section{In-band Full-Duplex MIMO Fundamentals}\label{sec:FD_MIMO}

%\textbf{Responsible authors: Besma  and then George and  Taneli}}
\subsection{Characterization of the Self-Interference Signal}
\begin{figure}[!t]
    \centering
    \includegraphics[width=7in]{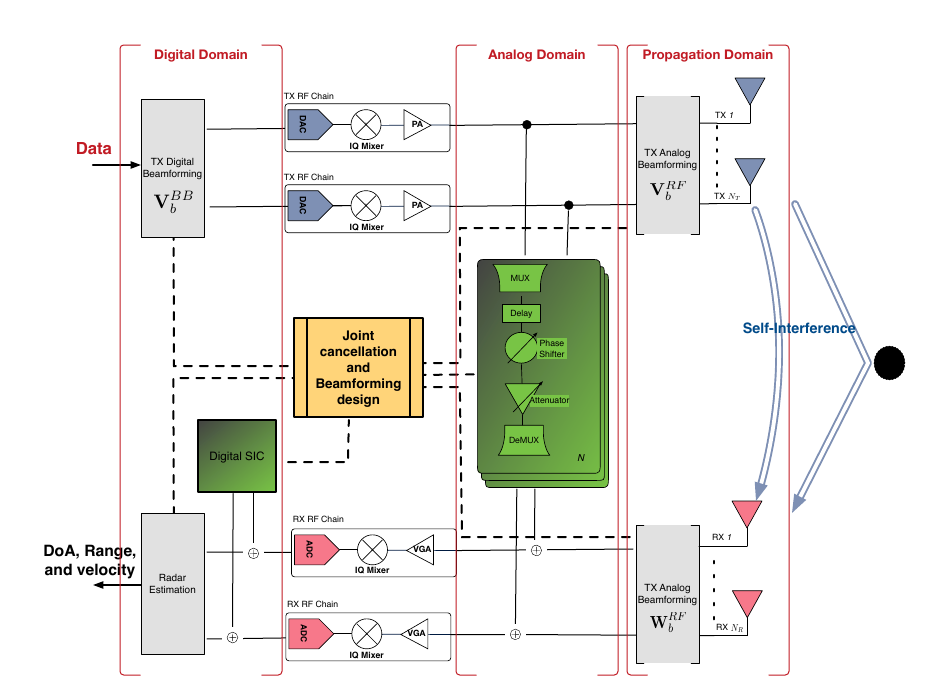}
    \caption{The architectural components of an IBFD $N_{\rm R}\times N_{\rm T}$ MIMO node $b$ realizing simultaneous DL data transmission and monostatic sensing.} 
    \label{fig:Framework}
\end{figure}

%In this section, we discuss the statistical characterization of the SI channels in IBFD MIMO. 
The SI channel between the RX and TX of an IBFD node comprises the direct path and any relevant multipath components. It can thus be modeled as a multipath channel with an exponential power delay profile, consisting of a quasi-static component due to the TX/RX antenna structures and a time-varying part related to the reflections from the surrounding environment~\cite{8276297}. In Fig.~\ref{fig:Framework}, an IBFD massive MIMO node $b$ including a partially-connected hybrid beamforming architecture at both the $N_{\rm T}$-antenna TX and $N_{\rm R}$-antenna RX sides that operates in the vicinity of $K$ targets is illustrated. Since, the DL signal from this node reflects back from the all passive radar targets present in the environment, the SI channel, $\mathbf{H}_{\rm SI}$, can be mathematically modeled as follows: 
 \begin{equation}\label{eq:1}
        \mathbf{H}_{\rm SI} = \underbrace{\sum_{k=1}^{K}\alpha_k e^{j2\pi \left(T_sf_{D,k} - \tau_k f_c\right)}\mathbf{a}_{N_{\rm R}}(\theta_k)\mathbf{a}_{N_{\rm T}}^\text{H}(\theta_k)}_{ \triangleq\mathbf{H}_{\rm radar}\text{, includes sensing information}} + \underbrace{\mathbf{H}_{b,b}}_{\text{ Direct SI channel between TX and RX }},
\end{equation}
where $\alpha_k$ (with $k=1,2,\ldots,K$) is the complex amplitude of the echo received from the $k$-th  target (including path-loss and radar cross section), $f_c$ is the carrier frequency, $T_s $ is the symbol duration,  and $\theta_k$ is the angular direction of the $k$-th target. 
The two-way propagation delay $\tau_k$ due to the range of the target translates to a phase shift across the frequency and the Doppler shift $f_{D,k}$, due to the moving velocity, translates to a phase shift across symbol duration. $\mathbf{H}_{b,b} \in \mathbb{C}^{N_{\rm R}\times N_{\rm T}}$ is the direct SI channel path between the node's TX and RX. In addition, $\mathbf{H}_{\rm SI}$ varies with antenna structure as $\mathbf{a}_{N_{\rm R}}(\cdot)$ and $\mathbf{a}_{N_{\rm T}}(\cdot)$ denote the mappings from the angle $\theta_k$ to the steering vectors of the RX and TX antennas, respectively. 
This SI characterization indicates that sensing information is completely contained in the SI signal. Note that, in IBFD communications, the SI signal has to be eliminated completely. 

In IBFD systems deployed for simultaneous DL and UL communications, the SI signal needs to be cancelled since it contaminates the UL signal of interest. However, as shown in \eqref{eq:1}, this signal includes target sensing information. To this end, to enable FD-ISAC, it is necessary to admit reflections from the surrounding objects and avoid canceling the respective useful echo signals, while ensuring sufficient mitigation of the direct SI channel between RX and TX.

\subsection{From Self-Interference Cancellation to Self-Interference Management}
 \label{SIC}
%ADD GEORGE TEXT HERE???

%Our exposition on self-interference and methods for its suppression are deliberately concise as there are excellent reviews and tutorials available on the topic [6], [15]–[17]. Instead, in this paper, we dive deeper into 

% The challenge of adopting  in-band full-duplex (FD) approach has always been self-interference (SI). The transmitted signal is also received and can be more than 100 dB stronger than the signal coming from the intended transmitter farther away.  Several self-interference cancellation have been proposed.  In this paper,  we will review the methods that were developed to address the special needs of FD-ISAC.  
%\subsubsection{Solution approaches: Propagation, Beam-forming, Analog and Digital}
For successful FD-ISAC systems, it is important to suppress the {\it direct} SI channel $\mathbf{H}_{b,b}$ to ensure sufficient TX/RX isolation at the FD node's RX side, while  the reflections from
the targets must be preserved, i.e., the channel $\mathbf{H}_{\rm radar}$. %This section  outlines  the fundamental characteristics of self-interference cancellation.  S
In IBFD MIMO transceivers, SI removal usually happens in three steps that can be distinguished based on their domain of operation (propagation, analog and digita domains), as shown in Fig.~\ref{fig:Framework}.  When various domain SI suppression mechanisms are effectively implemented in the ISAC context, FD-ISAC communication can achieve success.

\subsubsection{Propagation Domain --- Beamforming} This decreases SI power due to propagation losses, antenna element directionality, and/or beamforming usually achieved by MIMO. The beamforming controls  the amplitude and phase coefficients of each antenna element to form beams, to maximize or minimize transmit/receive signal at specific points. In particular, the FD-ISAC transmit beamformer is optimized such that the target channel gain gets maximized while SI power leakage gets minimized. The receive beamformer/combiner rejects the remaining SI while
providing a high beamforming gain in the direction
of targets. Beamforming can be: 1) digital, where each antenna is connected to a dedicated Radio-Frequency (RF) chain providing maximum flexibility, 2) analog, where all antenna elements are connected to {\it one} RF chain for power efficiency,  or 3) hybrid,  where the reduced number of RF chain offers a trade-off between flexibility of beam-patterns and power cost. To highlight the differences between the FD-ISAC and IBFD beamformer/combiner, we depict the optimized analog TX beamformer $\mathbf{V}_b^{\rm RF}$ and analog RX combiner $\mathbf{W}_b^{\rm RF}$  for FD-ISAC in Figure \ref{beam}. Note that TX gains at both sensing  and communication directions are maximized.  While the RX gain at the communication directions are attenuated. This marks a significant difference from IBFD MIMO SI cancellation, where the  TX and RX gains are maximized only in communication directions. 
 
\subsubsection{Analog Domain} The FD node can estimate the SI signal and subtract it from the received signal. Since its RX part {\it knows} the signal it is
transmitting, it can extract its own information after correcting for the channel effect. The analog canceller usually consists of few (but not many) taps, each one including a line of fixed delay, variable phase shifter, and attenuator~\cite{FD_MIMO_Arch}. The taps are then connected via simple multiplexers (MUXs)/demultiplexers (DEMUXs) for efficient signal routing among the TX/RX RF chains. Two such reduced complexity hardware architectures for the analog canceler in IBFD MIMO were presented in~\cite{FD_MIMO_Arch}. The one architecture was based on analog taps and the other on Auxiliary (AUX) TXs. In contrast to the available analog cancellation architectures, the values for each tap or each AUX TX and the configuration of the multiplexers were jointly designed with the digital transceiver beamforming filters according to desired performance objectives. The former tap-based analog canceler architecture for SI cancellation was extended to the wideband case in~\cite{WB_FD_MIMO_TWC2022}. % where also TX IQ imbalances and power amplification nonlinearities were handled via a joint optimization of Analog and Digital (A/D) SI cancellation together with the TX and receiver (RX) digital beamforming.
Other hardware impairments for IBFD MIMO together with imperfect channel knowledge were studied in~\cite{Islam2019unified,FD_MIMO_Impairments}. It is noted that, for lower complexity analog SI cancellation, the focus can be on its line-of-sight component. After this cancellation, the residual SI signal needs to satisfy the RF saturation constraint at each reception unit.

\subsubsection{Digital Domain} This is similar to the analog domain, but performed on digital signals, i.e., after the Variable Gain Amplifier (VGA) and ADC. After beamforming and analog cancellation, the residual SI includes TX and RX in-phase and quadrature imbalances, nonlinear Power-Amplifier (PA) distortions, and the RX's noise figure \cite{8761524}. These nonlinear distortions  can be cancelled using a combination of Maximum Likelihood (ML) methods, such as deep neural networks and transfer learning, as well as traditional model-based signal processing approaches\cite{9552213,9732685}. In the proposed FD-ISAC systems, the challenge is to distinguish between the echoes of targets and the direct SI signal which are coupling in the presence of noise. We expect that ML-based approaches for signal classification, such as the independent component analysis algorithm, can be beneficial due to the independent statistical characteristics of the two kinds of signals, i.e., moving targets versus static close reflectors. 
%A recent example can be found in [75?] where  the compressed sensing approach is employed for joint parameter estimation andsymbol demodulation.  
%We will furthermore improve the spectral efficiency by minimizing the amount of \emph{on-line} training by decoupling the time-varying  target location from the data estimation  or by utilizing \emph{transfer learning}, which combines elements of both off-line and on-line training and accumulates the information across the time intervals.
%It is expected that by using advanced ML based techniques, the sensing accuracy for FD-ISAC can be improved as in \cite{10024760}. 
We believe that deep learning approach can {\it track} the
angle parameters more accurately. 

%ADD NL HERE. 

%As the computational power of semiconductors has been growing exponentially as per Moore's law, cancellation in the digital domain is preferred to the analog domain. 
%\end{enumerate}

\subsubsection{Unified design for Massive MIMO} As the number of antenna elements increases, the interference signal leaking from the TX of the IBFD radio to its RX becomes more severe, thus, SI cancellation becomes more complex. To simplify the algorithmic design, reduce costs, and/or improve SI suppression, we can implement a {\it unified} IBFD MIMO architecture comprising A/D TX/RX beamforming, as well as A/D SI cancellation, which can be jointly designed for various performance objectives and complexity requirements~\cite{AlexandropoulosVTM_2023}. 
%We next describe a unified hybrid IBFD massive MIMO architecture which can be jointly optimized for various performance objectives and complexity requirements; 
This approach distributes the complexity where it is needed across the aforedescribed domains. 

\section{FD-Enabled Simultaneous Communications and Sensing}\label{sec:Sensing_FD_MIMO}
%\section{Monostatic Sensing with In-Band Full-Duplex MIMO}\label{sec:Sensing_FD_MIMO}
The MIMO FD-ISAC node illustrated in Fig.~\ref{fig:Framework} includes hybrid A/D TX/RX beamforming and A/D SI cancellation~\cite{AlexandropoulosVTM_2023}. This architecture is capable to implement {\it simultaneous} sensing, in a monostatic fashion, and DL data transmission. Assume, for example, that it is deployed to serve one or more users in the DL direction, while, in parallel, it receives and processes the reflections of the DL signals by one or more radar targets within its vicinity, with the goal to estimate their parameters. The A/D TX/RX beamformers and the A/D SI cancellation units can be jointly optimized to meet ISAC-oriented objectives, as will be presented in the sequel. For example, an FD-based ISAC system may be designed to optimize DL communication performance, while guaranteeing a sensing-based performance metric, or vice versa. It can be also designed to jointly optimize sensing and communications, guaranteeing in parallel a certain pair of sensing and communications performance. 
In the next section, we will study the feasibility of FD-ISAC by evaluating its sensing range in the case of a mmWave massive MIMO setting.   %where we focus 

%The DL signals are reflected by the multiple radar targets distributed within the communication environment, which are received and processed at the RX of BS node for radar targets’ parameter estimation enabling integrated sensing and communication. The BS is equipped with a transmit antenna array  realizing downlink analog beamforming and  digitally controlled receive antenna elements used for sensing.  A joint optimization algorithm for designing the A/D transmit and receive beamformers as well as the Self-Interference (SI) cancellation with the objective to maximize the achievable downlink and the accuracy of the radar target sensing performance.

%\GAadd{\subsection{Feasibility Study}\label{sec:Sensing_feasibility}
%\textbf{Responsible author: Atiq}\\
%Here we need to provide some numerical evidence using link budget and radar cross section; the dependence on the frequency and the beamforming gains needs to be showcased somehow to motivation the next subsection. We can add relevant performance results here.}\\
\subsection{Sensing Range Case Study}\label{sec:Sensing_feasibility}
\begin{figure}[!t]
    \centering
    \includegraphics[width=5in]{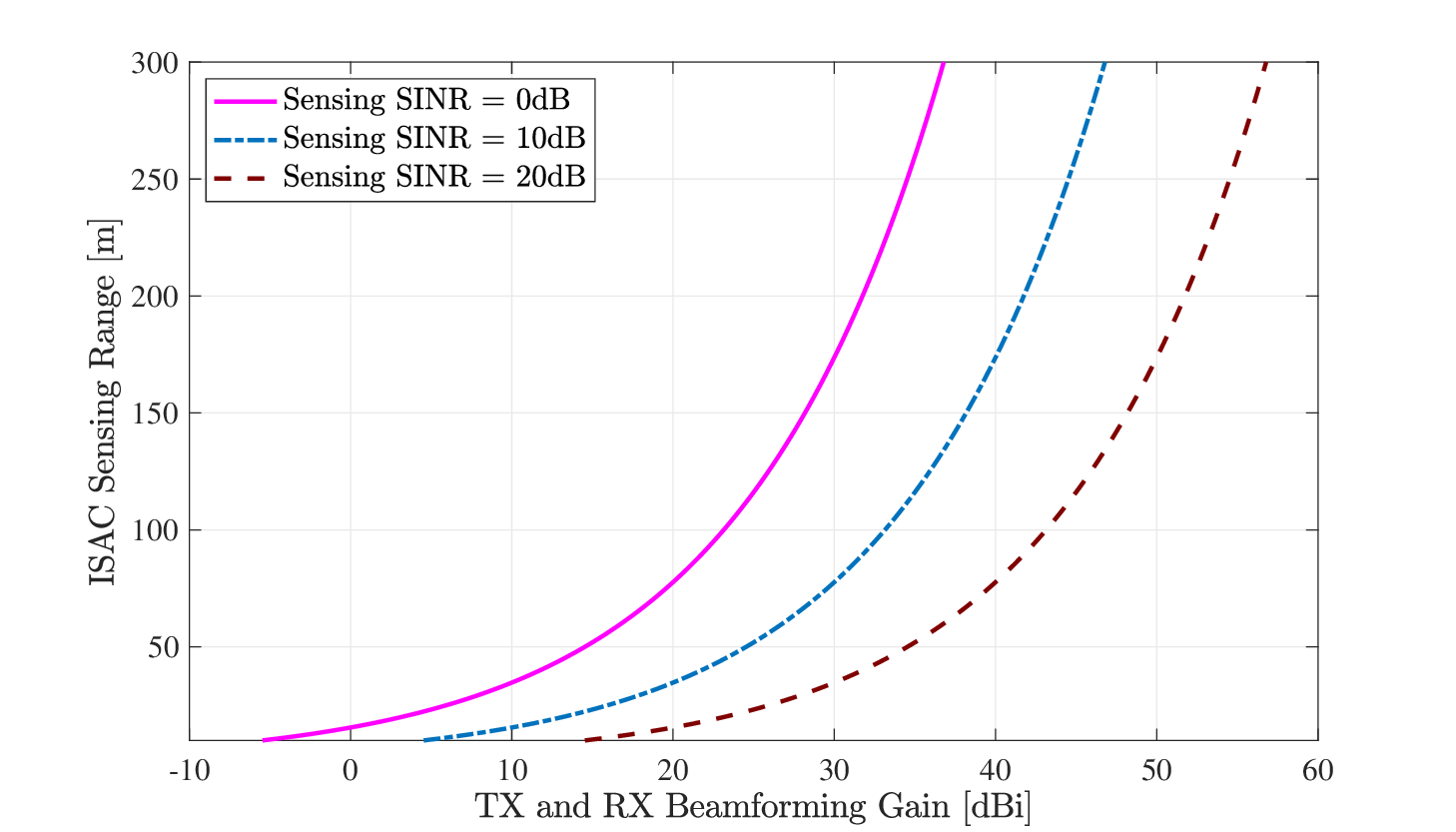}
    \caption{Sensing range of a $?\times?$ MIMO FD-ISAC system at $?$GHz with respect to the TX/RX beamforming gain for different SINR levels, $n_{p,k}=2.86$, $\sigma_{\rm{rcs,k}}=20$dBsm, and $\sigma_{s,k}=20$dB.} 
    \label{fig:sensing_range}
\end{figure}

In a mmWave massive MIMO FD-ISAC system, it is important to carefully adjust the transmit power at the FD node as well as the TX/RX beamforming gain to maintain stable sensing and communication links. It is, however, challenging to obtain precise sensing performance of the radar targets, since the received signal at the node's RX unit travels twice the sensing range, given the monostatic nature of the system. Considering the high path-loss of mmWave channels, it is of paramount importance to design A/D TX/RX beamformers for our FD-ISAC systems with high beamforming gain to achieve the required a certain level of Signal-to-Interference-plus-Noise-Ratio (SINR) at the RX. Starting from \eqref{eq:1}, the complex amplitude $\alpha_k$ of each $k$-th target can be defined as follows:
\begin{equation}
    \alpha_k = \frac{\lambda_c^2 \sigma_{\rm{rcs,k}}}{(4\pi)^2 d_k^{n_{p,k}} \sigma_{s,k}},
\end{equation}
where $d_k$, $n_{p,k}$, $\sigma_{\rm{rcs,k}}$, and $\sigma_{s,k}$ represent respectively the range, path-loss exponent, Radar Cross Section (RCS), and the combined small-scale fading and shadowing loss of the $k$-th target with $\lambda_c$ denoting the operating wavelength. Now we consider a practical mmWave massive MIMO BS operating at $28$GHz frequency with a bandwidth of $500$MHz and a transmit power of $30$dBm communicating with a DL user, while detecting multiple targets included in the sensing environment. In Fig.~\ref{fig:sensing_range}, we depict the sensing range of a $128\times128$ MIMO FD-ISAC system with respect to the TX/RX beamforming gain required to achieve different SINR levels, considering a line-of-sight mmWave channel at $28$GHz with path-loss exponent of $n_{p,k} = 2.86$ and $20$dB of combined small- and large-scale fading loss. We have assumed an automobile as the sensing target considering an RCS of $100m^2$ or $20$dBsm. As shown in the figure, approximately $40$dBi of TX/RX beamforming gain is required to successfully detect a radar target at the range of $150$m with a $10$dB SINR value.

%\GAadd{\subsection{Beamforming Design}
% \textbf{Responsible author: George}\\Here we need an example design that thies to tackle the latter problem (i.e., beamforming for optimizing sensing performance). We can add relevant performance results here.
\subsection{Beamforming Design}
We now consider the MIMO FD-ISAC node in Fig.~\ref{fig:Framework} in the case that wishes to communicate with a $N_{\rm ue}$-antenna user in the DL direction, while simultaneously sensing parameters related to a target placed in its vicinity. It is assumed that the $N_{\rm T}$ TX and $N_{\rm RX}$ RX antenna elements are connected with $N_{\rm T}^{\rm RF}$ TX and $N_{\rm R}^{\rm RF}$ RX RF chains, respectively; this implies the general case of hybrid A/D beamforming. The node's TX unit makes use of the $N_{\rm T}^{\rm RF}\times d$ precoding matrix $\mathbf{V}_b^{\rm BB}$ and the $N_{\rm T}\times N_{\rm T}^{\rm RF}$ analog beamforming matrix $\mathbf{V}_b^{\rm RF}$ to process its $d$-element symbol vector $\mathbf{x}$ ($d\leq\min\{N_{\rm ue},N_T^{\rm RF}\}$) before transmission, while the signal received at its RX unit can be processed in the analog domain via the $N_{\rm R}^{\rm RF} \times N_{\rm R}$ combining matrix $\mathbf{W}_b^{\rm RF}$, and, in the digital domain, via the $d \times N_{\rm R}^{\rm RF}$ combining matrix $\mathbf{W}_b^{\rm BB}$. The DL transmission obeys the node's total transmit power budget $P$, i.e., $\mathbb{E}\{ \| {{\mathbf{V}}_b^{{\text{RF}}}}\mathbf{V}_b^{\rm BB}\mathbf{x}\|^2\}\leq P$. Finally, to deal with the SI signal, the considered the node deploys the $N_{\rm R}\times N_{\rm T}$ analog- and digital-domain SIC matrices $\mathbf{A}$ and $\mathbf{D}$, respectively. 
%{\mathbb{E}}\left\{ {{{\left\| {{\mathbf{V}}_b^{{\text{RF}}}{\mathbf{V}}_b^{{\text{BB}}}} \right\|}^2}} \right\} \leq 

We next focus on the joint optimization of the latter MIMO FD-ISAC system parameters to enable simultaneous DL data communications and target parameter estimation in the UL, the latter via the received reflections of the DL signal (and not via dedicated pilots used in conventional non-FD-based monostatic radar systems). We particularly consider the following optimization formulation as an example (excluding the optimization of the DL user parameters or equivalently assuming optimum combining at this receiving user):
\begin{equation*}\label{eq:optim}
\begin{split}
  \mathcal{OP}:\quad &\max_{\mathbf{A},\mathbf{D},\mathbf{V}_b^{\rm RF},\mathbf{V}_b^{\rm BB},\mathbf{W}_b^{\rm RF},\mathbf{W}_b^{\rm BB}} \mathcal{C}\left(\mathbf{H}_{\rm DL},\mathbf{V}_b^{\rm RF},\mathbf{V}_b^{\rm BB} \right)
	\\& \textrm{s.t.}~~{\rm Tr}\left\{\mathbf{V}_b^{\rm RF}\mathbf{V}_b^{\rm BB}\left(\mathbf{V}_b^{\rm BB}\right)^{\rm H}\left(\mathbf{V}_b^{\rm RF}\right)^{\rm H}\right\}\leq P\,\,\,({\rm C1}),\\& \hspace{0.72cm}{\rm Hardware\,\,constraints\,\,on\,\,the\,\,structure\,\,of}\,\,\mathbf{A}\,\,\,({\rm C2}), \\& \hspace{0.72cm}f\left(\left(\mathbf{W}_b^{\rm RF}\right)^{\rm H}(\mathbf{H}_{\rm SI}+\mathbf{A})\mathbf{V}_b^{\rm RF}\mathbf{V}_b^{\rm BB}\mathbf{x}\right) \leq \boldsymbol{\lambda_{\rm SIC}}\quad({\rm C3}),\\& \hspace{0.72cm}\mathcal{S}\left(\mathbf{W}_b^{\rm RF},\mathbf{W}_b^{\rm BB},\mathbf{H}_{b,b},\mathbf{H}_{\rm radar},\mathbf{A},\mathbf{D},\mathbf{V}_b^{\rm RF},\mathbf{V}_b^{\rm BB}\right) \geq \lambda_{\rm S} \quad({\rm C4}),
\end{split}
\end{equation*}
where $\mathbf{H}_{\rm DL}$ represents the $N_{\rm ue}\times N_{\rm T}$ represents the DL channel matrix and the remaining channel matrices $\mathbf{H}_{\rm SI}$, $\mathbf{H}_{b,b}$, and $\mathbf{H}_{\rm radar}$ are defined in \eqref{eq:1}. In this example, yet parametric, MIMO FD-ISAC formulation, the design objective is the maximization of a DL communication metric $\mathcal{C}(\cdot)$ (e.g., the achievable DL rate), which depends on the digital and analog beamformers $\mathbf{V}_b^{\rm BB}$ and $\mathbf{V}_b^{\rm RF}$, respectively, while meeting a target parameter estimation objective $\mathcal{S}(\cdot)$ that relates to the gain of the received signal reflection from the target (e.g., the received signal-to-noise ratio~\cite{ISAC2022} or the opposite of the estimation Cram\'{e}r-Rao bound~\cite{RIS_ISAC}). As shown in constraint $({\rm C4})$, the latter gain depends on the A/D SIC matrices $\mathbf{A}$ and $\mathbf{D}$, the TX/RX beamformers $\mathbf{V}_b^{\rm BB}$, $\mathbf{V}_b^{\rm RF}$, $\mathbf{W}_b^{\rm BB}$, and $\mathbf{W}_b^{\rm RF}$, as well as the channel matrices $\mathbf{H}_{b,b}$ and $\mathbf{H}_{\rm radar}$, and $\lambda_{\rm S}$ represents a threshold indicating the imposed desired sensing estimation accuracy. As for the other constraints in $\mathcal{OP}$, $({\rm C1})$ describes the total transmit power budget at the MIMO FD-ISAC node and $({\rm C2})$ relates to the hardware capabilities and limitations of the analog canceller (see~\cite{FD_MIMO_Arch} for different forms of the analog cancellation matrix $\mathbf{A}$). Finally, constraint $({\rm C3})$ includes the general vector function $f(\cdot)$ of the instantaneous residual SI appearing at the node's RX antenna elements after analog SIC and before the RX RF chains, which needs to be upper bounded by the respective $N$ values in the vector $\boldsymbol{\lambda_{\rm SIC}}$ to avoid the saturation of the ADCs at the latter RF chains. It is noted that the example ISAC framework in $\mathcal{OP}$ aims to optimize a communication metric while ensuring that a specific sensing performance threshold is met. Alternatively, those metrics could swap positions on $\mathcal{OP}$'s objective and constraints, i.e., focus on optimizing a sensing metric while guaranteeing a required communication level, or they can be both incorporated in the objective and have dedicated thresholds as constraints.

%\BSadd{Radar beamforming -- Taneli}
In MIMO FD-ISAC, the analog beamformer $\mathbf{V}_b^{\rm RF}$ and combiner $\mathbf{W}_b^{\rm RF}$ are common for both the radar and communication operations. However, digital beamforming in MIMO radar is somewhat different from that in communications and communication-centric ISAC. In particular, in MIMO radar, digital beamforming is applied in an indirect way compared to communications, by embedding the spatial processing into the transmitted waveform from the multiple antenna. This radar waveform is fed directly to the TX RF chains and the digital vector with the received signals at the outputs of the RX RF chains is used directly for sensing before another copy is processed via $\mathbf{W}_b^{\rm BB}$ for detecting the UL communication symbols. This indicates that the spatial SI suppression becomes coupled with the MIMO waveform design, instead of being solely a joint cancellation and beamforming task as in conventional FD MIMO systems~\cite{FD_MIMO_Arch}.
%Beamforming in MIMO radar operation is somewhat different from that in communications and communication-centric ISAC, while the analog beamforming by $\mathbf{V}_b^{\rm RF}$ and $\mathbf{W}_b^{\rm RF}$ is applied in both radar and communications. Digital beamforming is applied in a MIMO radar in an indirect way compared to communications by embedding the spatial processing into the multi-antenna waveform for transmission. The MIMO radar waveform is fed directly to the TX RF chains and digital vector signal from the RX RF chains processed directly by MIMO radar RX chain before another copy is fed through $\mathbf{W}_b^{\rm BB}$ to UL communications RX chain. Due to this reason, the spatial SI suppression becomes coupled with MIMO waveform design, instead of being only a beamforming task.

\subsection{Waveform Design}
%\textbf{Responsible author: Taneli}\\Wideband case, extrapolation, OFMA or OTFS. We can add relevant performance results here.
The transmitted waveform in FD-ISAC systems realizing simultaneous communications and sensing necessitates multi-objective optimization that addresses the trade-off between the two targeted functionalities. The existence of SI may affect both communications and sensing directly or at least indirectly as a compromise for the one functionality to reduce the SI effect on the other. The waveform can be optimized in time, frequency, and space as well as, ideally, in combinations of all of them, which, however, contributes in increased complexity. In this way, the symbol vector $\mathbf{x}$ becomes a two-dimensional matrix (antennas $\times$ time) or even a three-dimensional tensor (antennas $\times$ time $\times$ frequency) constituting the digital waveform. The former is obtained from the latter by splitting the time dimension in intervals and applying the discrete Fourier transform; this is the case in the Orthogonal Frequency-Division Multiplexing (OFDM) scheme.

%The transmitted waveform design for IBFD MIMO systems that implement simultaneous communications and sensing requires multi-objective optimization and addressing the trade-off between communications and radar performance, where the self-interference may affect both functions directly and at least indirectly by as a compromise for one to reduce the effect of self-interference on the other. The waveform can be optimized in time, frequency and space as well as ideally in combinations all of them but the complexity, which makes the symbol vector $\mathbf{x}$ actually a two-dimensional matrix (antennas $\times$ time) or a three-dimensional tensor (antennas $\times$ time $\times$ frequency) for the digital waveform. The former is obtained from the latter by splitting the time dimension in intervals and applying discrete Fourier transform, i.e., Orthogonal Frequency-Division Multiplexing (OFDM).

Applying OFDM makes the MIMO waveform design more tractable. In
\cite{Liyanaarachchi2023waveform}, the $5$-th Generation New Radio (5G NR) OFDM waveform was optimized for joint communications and sensing with FD base stations. The proposed system maintained multiple spatial streams for communication links over frequency-selective non-line-of-sight channels, while simultaneously transmits separate spatial radar streams to different sensing directions. The received reflections from the environment due to all transmitted streams were used for sensing employing radar processing techniques. The presented design also optimized the A/D TX precoding and analog RX combining (i.e., $\mathbf{V}_b^{\rm BB}$, $\mathbf{V}_b^{\rm RF}$, and $\mathbf{W}_b^{\rm RF}$) so that inter- and intra-user interference as well as radar–communications interference are also canceled. This allowed to improve the detection probability and estimation errors of sensing via controlling the  information transmission rate through the communication objective.

%-------------------------------------------------------------------------
%	Section III: Monostatic Sensing with FD MIMO
%-------------------------------------------------------------------------
\begin{table}[tbp]
    \caption{Simulation Parameters.}
    \label{tab: sim_param}
    \centering
    % \footnotesize
    \begin{tabular}{|c|c||c|c|}
         \hline
          \textbf{Parameter}  &  \textbf{Value}  &  \textbf{Parameter}  &  \textbf{Value}  \\
         \hline
          Frequency  & $28$GHz  &  Number of TX/RX RF chains at node $b$ &  $8$  \\
          Bandwidth  & $500$MHz  &  Number of ULA antenna elements at node $b$ &  $16$ \\
          Number of OFDM symbols  &  $14$  &  Number of RX RF chains at node $u$ &  $4$ \\
          Subcarrier spacing  &  $120$kHz  &  Pathloss exponent  &  $2.86$ dBm \\
          Active subcarrier  &  $792$  &  Combined small- and large scale fading loss &  $8$ dBm \\
          OFDM symbol duration & $T_s = 8.92\mu$s  &  RCS (automobile) &  $100m^2$ ($20$dBsm) \\
          Peak-to-Average Power Ratio & $10$dB & SI channel pathloss & $40$dB \\
          Noise floor of nodes $b$ and $u$  &  $-87$ dBm & SI Channel $\kappa$-factor & $35$dB\\
          Noise Figure of nodes $b$ and $u$ &  $7$ and $3$ dB & DFT codebook resolution & $5$-bit\\
         \hline
    \end{tabular}
\end{table}
%\section{Simultaneous Communications and Sensing --- Numerical examples }
\section{Numerical Results and discussion}
In this section, we present performance evaluation results for a massive MIMO FD-ISAC system operating at mmWave frequencies which communicates with a DL MIMO user through the 5G NR OFDM waveform while simultaneously sensing its surrounding environment that includes multiple passive targets.

\textbf{Simulation Parameters:} We consider the MIMO FD-ISAC node $b$ in Fig.~\ref{fig:Framework} with $N_{\rm R}=N_{\rm T}=128$ antennas and a user with $N_{\rm ue}=4$ antenna elements that realizes fully digital combining in the DL direction. Each TX/RX RF chain at the FD node is connected to a Uniform Linear Array (ULA) capable of generating directive beams via a DFT codebook; it thus implements hybrid A/D TX/RX beamforming. The detailed parameters of the simulation are presented in Table.~\ref{tab: sim_param}. The FD-ISAC system is considered in a vehicular application with the goal to sense $K=6$ automobiles distributed within its vicinity with DoAs $\theta_k\in[-90^{\circ}, 90^{\circ}]$ with $k=1,2,\ldots,6$. The range and relative velocity for each automotive radar target are selected randomly with a maximum range of $80$m and maximum velocity of $100$km/h (i.e., $27.7\rm{ms}^{-1}$). For the DL communication with node $u$, $2$ out of $6$ targets were chosen randomly. 

\begin{figure}[!t]
	\begin{center}
	\includegraphics[width=0.84\linewidth]{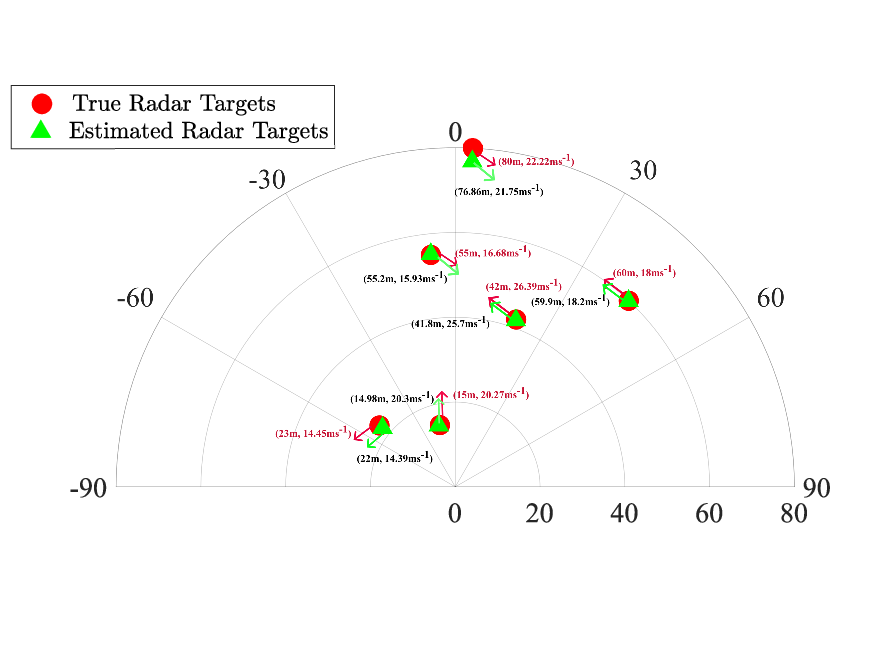}
	\caption{Sensing performance for $6$ automotive radar targets via the proposed $128\times128$ MIMO FD-ISAC system and optimization framework with $8$ TX and RF RF chains, each connected with a $16$-element ULA, and DL transmit power of $30$dBm.}\vspace{-0.2cm}
	\label{fig: sim_1}
	\end{center}
\end{figure}
\textbf{Complexity of the Analog SI Canceller:} To hope for efficient estimation of the targets' DoAs, the MIMO FD-ISAC node $b$ needs to efficiently cancel the strong SI signal. To this end, as also shown in Fig.~\ref{fig:Framework}, we consider both A/D SI cancellation to suppress this signal below the required radar SINR level. Since, for a practical mmWave setup with large number of antennas, it is infeasible to cancel the SI path from each of them, we implemented reduced complexity analog cancellation applied to the RX RF chain inputs, as shown in Fig.~\ref{fig:Framework}. Our aim was to utilize the smallest possible number of analog taps while avoiding RX RF chain saturation at node $b$; this is expressed in $\mathcal{OP}$'s constraint $({\rm C3})$. Note that, for the considered $128\times128$ MIMO FD-ISAC node with $8$ TX and $8$ RX RF chains, full-tap analog cancellation would need $64$ taps, whereas our analog canceller requires only $8$ analog taps (i.e., $12.5\%$ of the full-tap complexity). The residual SI was suppressed using a digital cancellation approach while maximizing the DL rate, as shown in $\mathcal{OP}$ with $\mathcal{C}(\cdot)$ being that rate and $\mathcal{S}(\cdot)$ the minimum SINR at the radar targets.

\textbf{MIMO FD-ISAC Optimization:} We have performed joint optimization of the beamforming and SI cancellation parameters of node $b$ (specifically, the A/D TX/RX beamformers $\mathbf{V}_b^{\rm RF}$, $\mathbf{W}_b^{\rm RF}$, $\mathbf{V}_b^{\rm BB}$, and $\mathbf{W}_b^{\rm BB}$, the MUX/DEMUXs in the $8$-tap analog canceller described by matrix $\mathbf{A}$, and the digital SI cancellation matrix $\mathbf{D}$), as described in $\mathcal{OP}$, to enable simultaneous DL communications to node $u$ and precise sensing operation at the RX of node $b$. For the latter functionality, we have considered as the sensing accuracy $\lambda_{\rm{S}}$ in $\mathcal{OP}$'s constraint $({\rm C4})$ the minimum SINR at the radar targets of $10$dB.

\begin{figure}[!t]
	\begin{center}
	\vspace{-5cm }\includegraphics[width=0.84\linewidth]{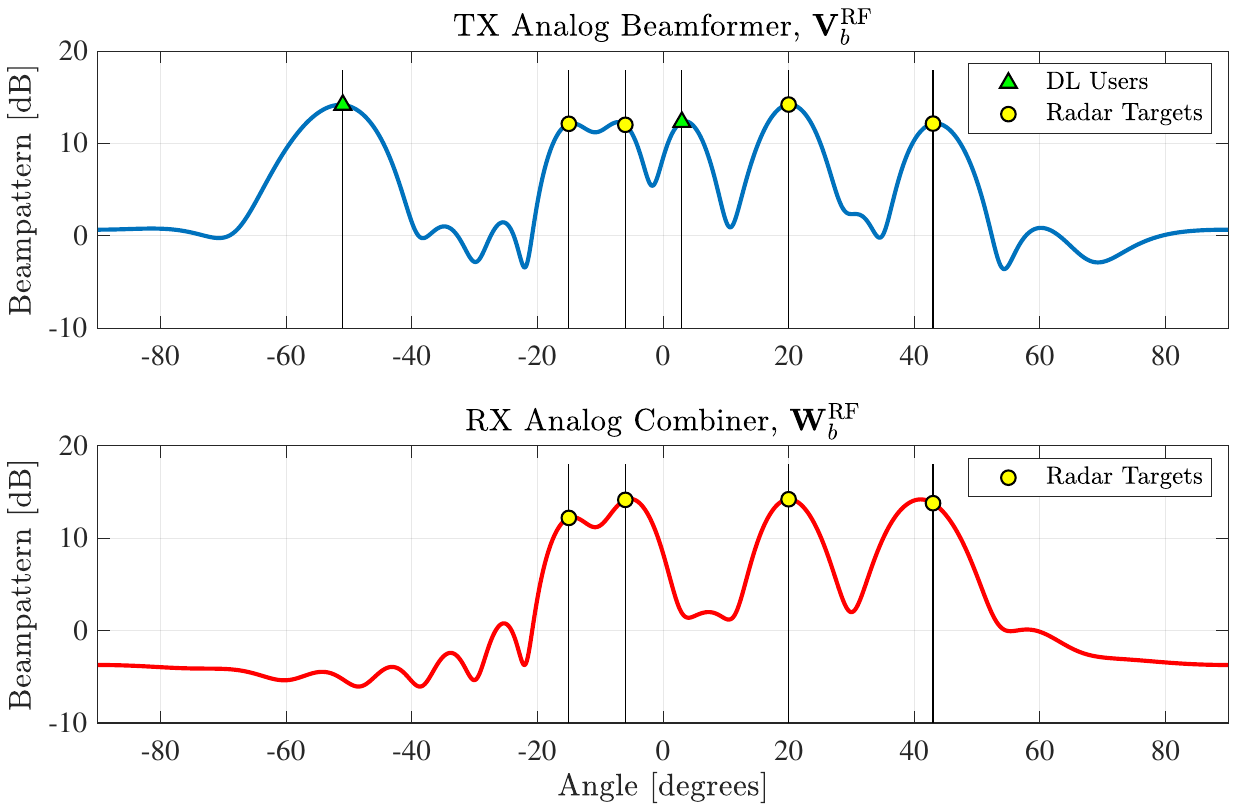}
 \vspace{-3cm}
	\caption{Analog TX beamformer $\mathbf{V}_b^{\rm RF}$ and Analog RX combiner $\mathbf{W}_b^{\rm RF}$ for $6$ automotive radar targets via the proposed $128\times128$ MIMO FD-ISAC system and optimization framework with $8$ TX and RF RF chains, each connected with a $16$-element ULA, and DL transmit power of $30$dBm.}\vspace{-0.2cm}
	\label{beam}
	\end{center}
\end{figure}

\begin{figure}[!t]
	\begin{center}
	\includegraphics[width=0.84\linewidth]{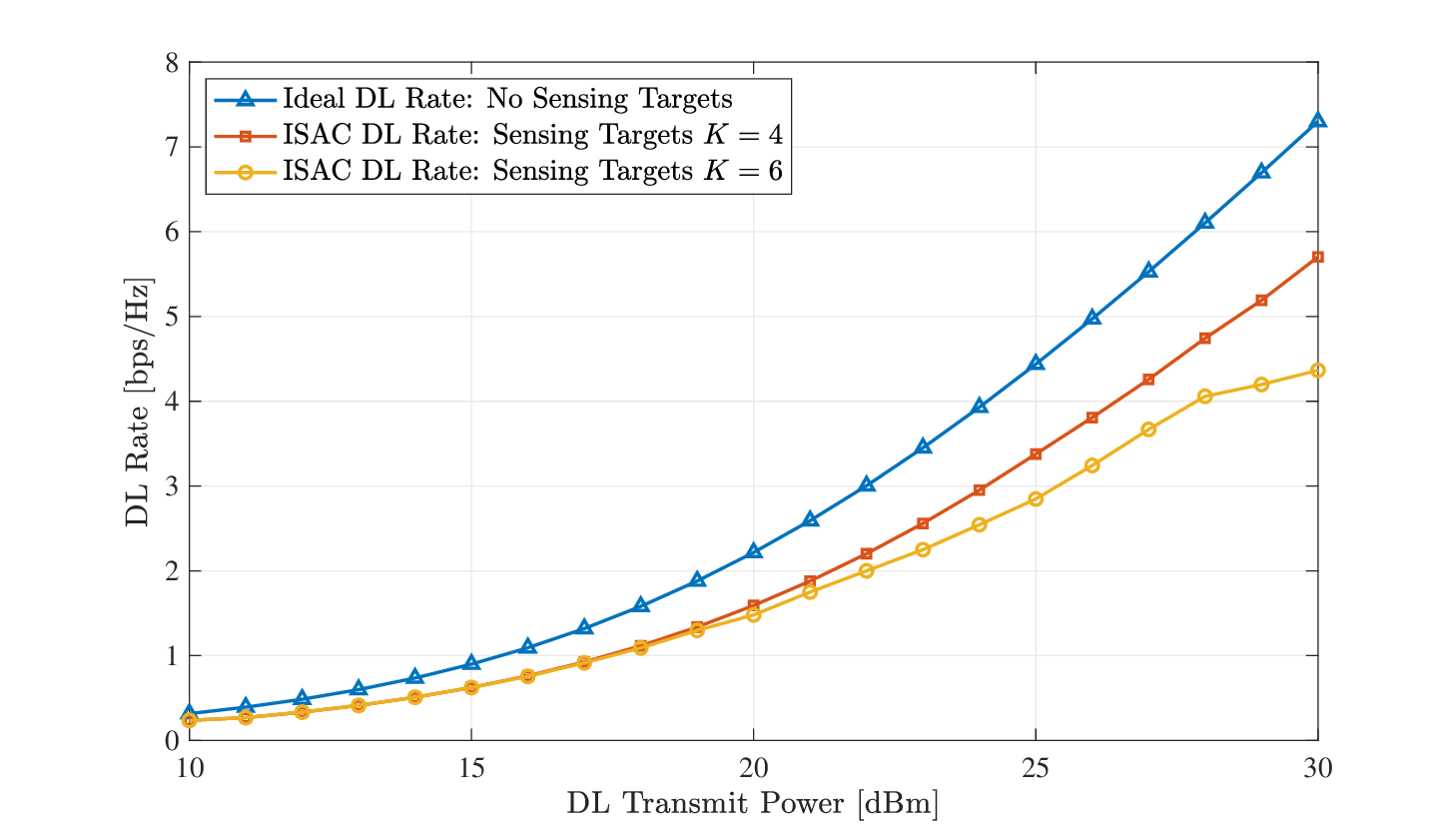}
	\caption{Achievable DL rate performance versus the transmit power $P$ in dBm for the MIMO FD-ISAC system and optimization framework in Fig.~\ref{fig: sim_1} considering different numbers $K$ of targets in the environment.}\vspace{-0.2cm}
	\label{fig: sim_2}
	\end{center}
\end{figure}

%\textbf{Automotive Radar Target Sensing performance:}
\textbf{Sensing performance:} The radar sensing performance of the proposed mmWave MIMO FD-ISAC system is illustrated in Fig.~\ref{fig: sim_1} for the transmit power $P=30$dBm. In particular, the estimated DoAs, ranges, and velocities of all $6$ targets are demonstrated in contrast to their true values. The true target DoAs are indicated with green triangles and black letters for the range and velocity values, where the red circles and letters refer to the respective estimations. It is evident that the sensing parameters of all  targets are successfully estimated with our joint optimization approach. Indicatively, for the targets placed up to $60$m, the range and velocity estimation error is $<1\%$. However, for the target located at a distance of $80$m, around $3$m range and $0.5\rm{ms^{-1}}$ velocity estimation errors occur, which is mainly attributed to the larger path-loss. We also plot the Analog TX beamformer $\mathbf{V}_b^{\rm RF}$ and Analog RX combiner $\mathbf{W}_b^{\rm RF}$ in Figure \ref{beam}.  The TX gains at both communications and sensing directions are almost similar, showing that all streams are transmitted to both sets of directions. The RX gains at sensing directions are maximized while those at the communication directions are attenuated. 

\textbf{Achievable Rate Performance:} The DL rate with the proposed MIMO FD-ISAC system and optimization framework in Fig.~\ref{fig: sim_1} for the case of a $4$-element DL user is plotted versus $P$ in Fig.~\ref{fig: sim_2} for different numbers $K$ of targets in the environment. As observed, the achievable rate reduces with increasing $K$ values, indicating that a more demanding sensing constraint $({\rm C4})$ penalizes the communication performance. Specifically, the considered optimization allocates more DL beamforming resources towards the targets for better sensing accuracy. For example, for $K=4$ radar targets, $1.34$bps/Hz more DL rate than the case of $K=6$ targets can be achieved.

\section{Future Research Directions}\label{sec:Directions}
As previously discussed, ISAC in the explict form of simultaneous data transmission and monostatic sensing operations can be efficiently realized with appropriately designed FD MIMO systems. In this section, we present some of the key open challenges and their resulting research directions with MIMO FD-SAC.

\emph{Theoretical foundation}: One of the key open areas ripe for research is the development of the theoretical foundations to understand the performance limits of supporting two functions simultaneously: communications and sensing. Past works have researched estimation-theoretic metrics for sensing in ISAC~\cite{Bliss:2014,Paul:2016,Li:2021} and information-theoretic approaches to ISAC~\cite{Chiriyath:2015a, 10147248}, and more recently a degrees-of-freedom perspective. The diversity of ideas in analyzing joint communications and sensing systems demonstrates the range of perspective in the choice of metrics, which in turn means there is much to understand in the relationships between metrics and how they impact performance trade-offs. Continuing from the existing studies to incorporate IBFD operation and the resulting SI opens major new research questions as this adds a new dimension to the interplay between communications and sensing.

\emph{FD-ISAC modeling}: The 3rd Generation Partnership Project (3GPP) Release $19$ recently launched channel modeling work for ISAC as well as new spectrum. The former activity aims to accurately model signal reflection and multi-path propagation, initially focusing on sensing of passive objects. MIMO FD-ISAC contributes the simultaneous transmission and reception feature, resulting in SI comprosing of possibly mixed near- and far-field channel terms. To this end,  accurate, while tractable, channel and signal models capturing different settings and conditions are required. The models need to capture both the dominant effects accurately, e.g., the SI channel, non-linearities, and noise statistics, which may differ from those characterized previously for communications-only scenarios.

\emph{Physical-layer design}: 
Given the channel, signal models and design objectives, there is significant room for innovation in the design of IBFD waveforms that can flexibly achieve different trade-off points incorporating the technical challenge of SI. For example, in~\cite{Mehrotra:2021a}, the authors derived a degree-of-freedom trade-off between wireless imaging and UL communications. Different points on the trade-off require different physical-layer designs. In addition, there exists a lot of room of MIMO FD-ISAC problem formulations and joint designs of all involved parameters (i.e., hybrid A/D TX/RX beamforming and A/D treatment of the SI signal). To this end, the trend towards extremely massive MIMO and beyond mmWave frequencies imposes challenges in the beamforming design, but also the opportunity to treat SI without dedicated cancellation units~\cite{FD_HMIMO_2023}.

\emph{Hardware platforms}: To validate the performance of advanced FD ISAC techniques at the mmWave frequency band, it is essential to develop dedicated hardware platform capable of hybrid A/D TX/RX beamforming with both analog and digital treatment of the SI signal. Very recently, a mmWave FD MIMO hardware prototype was presented in~\cite{yu2023realizing}, demonstrating the ability to suppress interference in the spatial, analog, and digital domains, thereby significantly reducing residual SI to the noise floor. Additionally, the authors in~\cite{pham2021joint} presented a mmWave half-duplex ISAC platform, showcasing its sensing and communication performance with multiple targets. However, there remains a need for a mmWave MIMO FD-ISAC hardware platform with simultaneous transmission and monostatic sensing capability.

\emph{Near-field operation}: The use of Reconfigurable Intelligent Surfaces (RISs)~\cite{FD_RIS_EUSIPCO2023} and metasurface-based antennas~\cite{FD_HMIMO_ICASSP2024} at mmWaves and beyond (i.e., terahertz) are particularly suitable for sensing, thus, complementing typical sensing systems. In the former work, it was shown that placing an RIS in the near-field region of a MIMO FD-ISAC node can boost its ISAC capability when appropriately designing its parameters jointly with that of node's. On the other hand, the latter work emphasized that a scalable MIMO FD-ISAC framework profits from near-field signal propagation. However, these works constitute preliminary investigations and mode sophisticated signal and channel models, new waveforms and related beamforming approaches will be needed. Especially, there is very limited work on MIMO FD-ISAC for near-field taking into account SI in both the waveform and beamforming designs.

\section{Conclusion}\label{sec:FD_MIMO}
In this article, we presented a generic MIMO FD-ISAC framework for spectrum-efficient simultaneous information transmission and monostatic sensing, elaborating upon its unique properties and associated SI challenges, as well as surveying state-of-the-art special cases. The framework's fundamental difference in treating SI with respect to conventional FD MIMO systems used for simultaneous DL and UL communications was discussed, and future research directions with MIMO FD-ISAC systems and signal processing were presented. We believe that FD's potential for ISAC will likely keep the research community occupied for many years to come, considering that there are a number of exciting open research topics.

%---------------------------------------------------------------------------------------
%	References
%---------------------------------------------------------------------------------------
%\balance
\bibliographystyle{IEEEtran}
\bibliography{IEEEabrv,ref,refsBesma}

\begin{thebibliography}{10}
\providecommand{\url}[1]{#1}
\csname url@samestyle\endcsname
\providecommand{\newblock}{\relax}
\providecommand{\bibinfo}[2]{#2}
\providecommand{\BIBentrySTDinterwordspacing}{\spaceskip=0pt\relax}
\providecommand{\BIBentryALTinterwordstretchfactor}{4}
\providecommand{\BIBentryALTinterwordspacing}{\spaceskip=\fontdimen2\font plus
\BIBentryALTinterwordstretchfactor\fontdimen3\font minus
  \fontdimen4\font\relax}
\providecommand{\BIBforeignlanguage}[2]{{%
\expandafter\ifx\csname l@#1\endcsname\relax
\typeout{** WARNING: IEEEtran.bst: No hyphenation pattern has been}%
\typeout{** loaded for the language `#1'. Using the pattern for}%
\typeout{** the default language instead.}%
\else
\language=\csname l@#1\endcsname
\fi
#2}}
\providecommand{\BIBdecl}{\relax}
\BIBdecl

\bibitem{3GPPTS22137}
3GPP, ``Ts 22.137 study on integrated sensing and communication, release 19,''
  3GPP SA1, Tech. Rep., 2023.

\bibitem{ETSIISG}
ETSI, ``Integration sensing and communication industry specification group,''
  ETSI, Tech. Rep., 2023.

\bibitem{FMC22}
F.~Liu, Y.~Cui, C.~Masouros, J.~Xu, T.~X. Han, Y.~C. Eldar, and S.~Buzzi,
  ``Integrated sensing and communications: Toward dual-functional wireless
  networks for {6G} and beyond,'' \emph{IEEE J. Sel. Areas Commun.}, vol.~40,
  no.~6, pp. 1728--1767, 2022.

\bibitem{AlexandropoulosVTM_2023}
G.~C. Alexandropoulos, M.~A. Islam, and B.~Smida, ``Full duplex massive {MIMO}
  architectures: {R}ecent advances, applications, and future directions,''
  \emph{IEEE Veh. Technol. Mag.}, vol.~17, no.~4, pp. 83--91, Dec. 2022.

\bibitem{10158711}
Z.~He, W.~Xu, H.~Shen, D.~W.~K. Ng, Y.~C. Eldar, and X.~You, ``Full-duplex
  communication for {ISAC}: Joint beamforming and power optimization,''
  \emph{IEEE Journal on Selected Areas in Communications}, vol.~41, no.~9, pp.
  2920--2936, 2023.

\bibitem{7756408}
D.~Korpi, M.~Heino, C.~Icheln, K.~Haneda, and M.~Valkama, ``Compact inband
  full-duplex relays with beyond 100 db self-interference suppression: Enabling
  techniques and field measurements,'' \emph{IEEE Transactions on Antennas and
  Propagation}, vol.~65, no.~2, pp. 960--965, 2017.

\bibitem{10463523}
B.~Smida, R.~Wichman, K.~E. Kolodziej, H.~A. Suraweera, T.~Riihonen, and
  A.~Sabharwal, ``In-band full-duplex: The physical layer,'' \emph{Proceedings
  of the IEEE}, pp. 1--30, 2024.

\bibitem{SmidaJSAC_2023}
B.~Smida, A.~Sabharwal, G.~Fodor, G.~C. Alexandropoulos, H.~A. Suraweera, and
  C.-B. Chae, ``Full-duplex wireless for {6G}: Progress brings new
  opportunities and challenges,'' \emph{IEEE J. Sel. Areas Commun.}, vol.~41,
  no.~9, pp. 2729--2750, Sep. 2023.

\bibitem{barneto2020beamforming}
C.~B. Barneto, S.~D. Liyanaarachchi, T.~Riihonen, M.~Heino, L.~Anttila, and
  M.~Valkama, ``Beamforming and waveform optimization for {OFDM}-based joint
  communications and sensing at mm-waves,'' in \emph{Proc. IEEE Asilomar
  Signals Sys. Comp. Conf.}, Nov. 2020, pp. 895--899.

\bibitem{liyanaarachchi2021joint}
S.~D. Liyanaarachchi, C.~B. Barneto, T.~Riihonen, M.~Heino, and M.~Valkama,
  ``Joint multi-user communication and {MIMO} radar through full-duplex hybrid
  beamforming,'' in \emph{Proc. IEEE Int. Symp. Joint Commun. \& Sensing
  (JC\&S)}, Feb. 2021, pp. 1--5.

\bibitem{Comms-Target_Tracking2022}
M.~A. Islam, G.~C. Alexandropoulos, and B.~Smida, ``Simultaneous multi-user
  {MIMO} communications and multi-target tracking with full duplex radios,'' in
  \emph{Proc. IEEE Global Commun. Conf. (GLOBECOM)}, Dec. 2022, pp. 1--6.

\bibitem{ISAC2022}
------, ``Integrated sensing and communication with millimeter wave full duplex
  hybrid beamforming,'' in \emph{Proc. IEEE Intl. Conf. Commun. (ICC)}, May
  2022, pp. 1--6.

\bibitem{FD_HMIMO_2023}
I.~Gavras, M.~A. Islam, B.~Smida, and G.~C. Alexandropoulos, ``Full duplex
  holographic {MIMO} for near-field integrated sensing and communications,'' in
  \emph{European Signal Proces. Conf. (EUSIPCO)}, Helsinki, Finland, Sep. 2023.

\bibitem{Asilomar_FD_HMIMO_2023}
M.~Talha, B.~Smida, G.~C. Alexandropoulos, and M.~A. Islam, ``Multi-target
  two-way integrated sensing and communications with full duplex mimo radios,''
  in \emph{Asilomar Signals, Systems, Comp. Conf.}, Pacific Grove, USA, Nov.
  2023.

\bibitem{Direction-Aided2020}
M.~A. Islam, G.~C. Alexandropoulos, and B.~Smida, ``Direction-assisted beam
  management in full duplex millimeter wave massive {MIMO} systems,'' in
  \emph{Proc. IEEE Global Commun. Conf. (GLOBECOM)}, Dec. 2021, pp. 1--6.

\bibitem{8276297}
F.~Chen, R.~Morawski, and T.~Le-Ngoc, ``Self-interference channel
  characterization for wideband 2 × 2 mimo full-duplex transceivers using
  dual-polarized antennas,'' \emph{IEEE Transactions on Antennas and
  Propagation}, vol.~66, no.~4, pp. 1967--1976, 2018.

\bibitem{FD_MIMO_Arch}
G.~C. Alexandropoulos, ``Low complexity full duplex {MIMO} systems: {A}nalog
  canceler architectures, beamforming design, and future directions,''
  \emph{ITU J. Future Evolving Technol.}, vol.~2, no.~2, pp. 1--19, Dec. 2021.

\bibitem{WB_FD_MIMO_TWC2022}
M.~A. Islam, G.~C. Alexandropoulos, and B.~Smida, ``Joint analog and digital
  transceiver design for wideband full duplex {MIMO} systems,'' \emph{IEEE
  Trans. Wireless Commun.}, vol.~21, no.~11, pp. 9729--9743, Nov. 2022.

\bibitem{Islam2019unified}
------, ``A unified beamforming and {A/D} self-interference cancellation design
  for full duplex {MIMO} radios,'' in \emph{Proc. {IEEE PIMRC}}, Sep. 2019.

\bibitem{FD_MIMO_Impairments}
H.~Iimori, G.~Abreu, and G.~C. Alexandropoulos, ``{MIMO} beamforming schemes
  for hybrid {SIC FD} radios with imperfect hardware and {CSI},'' \emph{IEEE
  Trans. Wireless Commun.}, vol.~18, no.~10, pp. 4816--4830, Oct. 2019.

\bibitem{8761524}
M.~A. Islam and B.~Smida, ``A comprehensive self-interference model for
  single-antenna full-duplex communication systems,'' in \emph{Proc. IEEE
  International Conference on Communications (ICC)}, May 2019, pp. 1--7.

\bibitem{9552213}
K.~Muranov, M.~A. Islam, B.~Smida, and N.~Devroye, ``On deep learning assisted
  self-interference estimation in a full-duplex relay link,'' \emph{IEEE
  Wireless Commun. Lett.}, vol.~10, no.~12, pp. 2762--2766, 2021.

\bibitem{9732685}
D.~H. Kong, Y.-S. Kil, and S.-H. Kim, ``Neural network aided digital
  self-interference cancellation for full-duplex communication over
  time-varying channels,'' \emph{IEEE Trans. Veh. Technol.}, vol.~71, no.~6,
  pp. 6201--6213, 2022.

\bibitem{RIS_ISAC}
S.~P. Chepuri, N.~Shlezinger, F.~Liu, G.~C. Alexandropoulos, S.~Buzzi, and
  Y.~C. Eldar, ``Integrated sensing and communications with reconfigurable
  intelligent surfaces,'' \emph{IEEE Signal Process. Mag.}, vol.~40, no.~6, p.
  41–62, Sep. 2023.

\bibitem{Liyanaarachchi2023waveform}
S.~D. Liyanaarachchi, T.~Riihonen, C.~B. Barneto, and M.~Valkama, ``Joint
  {MIMO} communications and sensing with hybrid beamforming architecture and
  {OFDM} waveform optimization,'' \emph{IEEE Transactions on Wireless
  Communications}, 2023.

\bibitem{Bliss:2014}
D.~W. Bliss, ``Cooperative radar and communications signaling: The estimation
  and information theory odd couple,'' in \emph{IEEE Radar Conference}, 2014.

\bibitem{Paul:2016}
B.~Paul, A.~R. Chiriyath, and D.~W. Bliss, ``Joint communications and radar
  performance bounds under continuous waveform optimization: The waveform
  awakens,'' in \emph{IEEE Radar Conference}, 2016.

\bibitem{Li:2021}
C.~Li, N.~Raymondi, B.~Xia, and A.~Sabharwal, ``"{Outer Bounds for a Joint
  Communicating Radar (Comm-Radar): Three-node Uplink}",'' \emph{IEEE
  Transactions on Communications}, November 2021.

\bibitem{Chiriyath:2015a}
A.~R. Chiriyath and D.~W. Bliss, ``Joint radar-communications performance
  bounds: Data versus estimation information rates,'' in \emph{Proc. IEEE
  Military Communications Conference (MILCOM 2015)}, 2015, pp. 1491--1496.

\bibitem{10147248}
Y.~Xiong, F.~Liu, Y.~Cui, W.~Yuan, T.~X. Han, and G.~Caire, ``On the
  fundamental tradeoff of integrated sensing and communications under
  {G}aussian channels,'' \emph{IEEE Transactions on Information Theory},
  vol.~69, no.~9, pp. 5723--5751, 2023.

\bibitem{Mehrotra:2021a}
N.~Mehrotra and A.~Sabharwal, ``Degrees of freedom analysis for
  multipath-assisted communication \& sensing,'' \emph{IEEE Journal on Selected
  Areas in Communications}, vol.~40, no.~6, pp. 1768--1779, Jun. 2022.

\bibitem{yu2023realizing}
B.~Yu, C.~Qian, J.~Lee, S.~Shao, Y.~Shen, W.~Pan, P.~Lin, Z.~Zhang, S.~Kim,
  S.~Hu \emph{et~al.}, ``Realizing high power full duplex in millimeter wave
  system: Design, prototype and results,'' \emph{IEEE Journal on Selected Areas
  in Communications}, vol.~41, no.~9, pp. 2893--2906, Jun. 2023.

\bibitem{pham2021joint}
T.~M. Pham, R.~Bomfin, A.~Nimr, A.~N. Barreto, P.~Sen, and G.~Fettweis, ``Joint
  communications and sensing experiments using mmwave platforms,'' in
  \emph{Proc. IEEE Workshop on Sig. Proc. Advan. in Wireless Commun. (SPAWC)},
  Lucca, Italy, Sep. 2021, pp. 501--505.

\bibitem{FD_RIS_EUSIPCO2023}
C.~K. Sheemar, G.~C. Alexandropoulos, D.~Slock, J.~Querol, and S.~Chatzinotas,
  ``Full-duplex-enabled joint communications and sensing with reconfigurable
  intelligent surfaces,'' in \emph{Proc. European Signal Process. Conf.
  (EUSIPCO)}, Helsinki, Finland, Sep. 2023, pp. 1--5.

\bibitem{FD_HMIMO_ICASSP2024}
I.~Gavras and G.~C. Alexandropoulos, ``Joint near-field target tracking and
  communications with full duplex holographic {MIMO},'' in \emph{Proc. IEEE
  Int. Conf. Acoustics, Speech, Signal Process. (ICASSP)}, Seoul, South Korea,
  Apr. 2024, pp. 1--5.

\end{thebibliography}
\end{document}